\documentclass[universe,article,accept,pdftex,moreauthors]{Definitions/mdpi} 
\firstpage{1} 
\pubvolume{1}
\issuenum{1}
\articlenumber{0}
\pubyear{2026}
\copyrightyear{2025}
\externaleditor{Firstname Lastname} 
\datereceived{9 November 2025} 
\daterevised{23 December 2025} 
\dateaccepted{26 December 2025} 
\datepublished{ } 
\hreflink{https://doi.org/} 

\Title{Analysis of 14 Years of X-Ray Emission From SN 2011DH}

\Author{{Elisa}
 {J.}
 Gao and Vikram {V.} 
 Dwarkadas {*} 
}

\AuthorNames{Elisa Gao, Vikram Dwarkadas}

\address[1]{
{Department} 
 of Astronomy and Astrophysics, University of Chicago, 5640 S Ellis Ave, Chicago, IL 60637, USA; elisagjy@uchicago.edu\\
}

\corres{\hangafter=1 \hangindent=1.05em \hspace{-0.82em}Correspondence: vikramvd@uchicago.edu; Tel.:  +1-773-834-4668}

\abstract{Ejecta from core-collapse supernovae interact with the circumstellar medium shed by the progenitor star, producing X-ray emission. Previous studies analyzed the X-ray spectrum of the Type~IIb supernova SN~2011dh up to $\sim$500~days after explosion. Long-term monitoring of X-ray emission provides valuable constraints on supernova evolution and progenitor systems, yet such studies remain rare for Type~IIb events due to limited data. Here we present the most comprehensive X-ray light curve of SN~2011dh to date, combining all available \textit{Chandra} and \textit{XMM-Newton} data with previously published and newly released \textit{Swift} observations, extending coverage to $\sim$5100~days. We measure a luminosity decline consistent with $L_X \propto t^{-0.74 \pm 0.04}$ and infer a mass-loss rate of $(1.0$–$2.2)\times10^{-6}~M_\odot~\mathrm{yr}^{-1}$ for $v_w = 10~\mathrm{km~s^{-1}}$, or $(2.0$–$4.4)\times10^{-6}~M_\odot~\mathrm{yr}^{-1}$ for $v_w = 20~\mathrm{km~s^{-1}}$. These estimates agree with earlier results, supporting the interpretation that the X-ray emission has been dominated by an adiabatic reverse shock.  The consistency of our late-time results with previous studies demonstrates that SN~2011dh has evolved steadily for nearly 14~years.
}


\begin{document}

\section{Introduction}

Core-collapse supernovae (CCSNe) result from the explosion of massive stars at the end of their lives. During their lifetime, these stars lose mass through stellar winds, creating a circumstellar medium (CSM) around the star. The material ejected in the supernova explosion is expanding at velocities $ > $ 5000 km s$^{-1}$. It collides with the CSM, {driving a shock} that can heat the CSM to temperatures $ \geq 10^7$ K. In order to slow down the ejecta, an  inward-propagating reverse shock is also generated, which expands into the higher density ejecta with a lower velocity, and consequently a lower temperature. The low temperatures seen in X-rays suggest that the reverse shock is the primary site of X-ray emission \citep{franssonetal96,chandra2009} in many SNe. The evolution of the SN shock waves strongly depends on the mass-loss history of the progenitor before explosion. {In a self-similar model \cite{1982}
  the properties of the gas behind the reverse shock can be related to those behind the forward shock \cite{cf17}, even if the X-ray emission arises from the reverse shock. Therefore, reverse shock emission also becomes a good measure of the circumstellar interaction. Thus, tracking the X-ray emission over time provides a powerful probe of the progenitor’s properties and late-stage behavior.}

Type IIb SNe are those that show a small amount of H in the optical spectrum initially. At later times this H may not be seen, and the SN resembles more a Type Ib SN without H. Ten Type IIb SNe have so far been detected in X-rays \citep{dwarkadas25}. The most extensively observed one is SN 1993J. When it was first detected, SN 1993J was the closest SN in the northern hemisphere in modern times, being located in M81 at a distance of only 3.6~Mpc \citep{freedman}. It remains the best-studied Type~IIb SN, with a wealth of data available over the entire wavelength spectrum from radio to X-ray. Its mass-loss rate is reasonably well constrained~\citep{ta,bjornsson25}. On the other hand, the nature of the circumstellar medium (CSM) created by the progenitor star has been debated for years, at both X-ray and radio wavelengths \citep{suzukietal93,sn95,franssonetal96, fb98,mdb01,weileretal07,bjornsson15,bjornsson25}. Its X-ray light curve extends at least up to 15~years after the explosion \citep{franssonetal96, ta, chandra2009, vposter}. The luminosity of SN~1993J appears to have decayed consistently until $\sim$3300~days, when a sudden drop in luminosity occurred. The exact cause of this change in evolution remains uncertain, with proposed explanations including a CSM with density modifications, perhaps due to a binary progenitor system. 

Unfortunately, other Type IIb SNe have not been investigated in similar detail in the X-ray regime. This is due both to their larger distance as well as the lack of continual follow-up.  Archival X-ray data, however, are available on many Type IIb SNe. SN 2011dh is a Type IIb SN that has been routinely monitored with the \textit{Chandra},
 \textit{XMM} and \textit{Swift} X-ray satellites. X-ray data are available from 2011 to late 2025, covering up to $\sim$5100~days after the explosion. X-ray analysis by \citet{maeda} considered the first 500~days after the explosion. Since then, no further work has been conducted on the X-ray light curve of SN 2011dh.  Given the large amount of archival X-ray data from various satellites, SN 2011dh is worth studying for the long-term evolution of the X-ray light curve, which can be compared to that of SN 1993J. Such an investigation could provide insight into Type~IIb supernova evolution in general, while adding one more well-studied example of an X-ray Type IIb SN besides SN 1993J. In addition, there are not many SNe that are well monitored for over a decade. SN 2011dh, as shown below, has an extensive series of observations, extending to $>$5000 days.

The first detection of SN~2011dh in M51 was reported by Amadee Riou on 2011 May 31.893 UT {(JD 2455713.393)}. It was subsequently identified as a Type~II supernova by~\cite{silver}. Early optical spectra analyzed by \cite{ar,marion} further classified it as a Type~IIb supernova---one that shows hydrogen in its early spectra but later evolves to become helium-dominated~\citep{fili}. 

Based on early spectroscopy, the progenitor of SN 2011dh was assumed to have a radius $\sim$$10^{11}$~$\mathrm{cm}$  \citep{soderberg}, making it more compact than the progenitor of the well-studied Type IIb supernova SN~1993J. The detection of a luminous star in archival pre-explosion \textit{Hubble Space Telescope} images, with a radius of $\sim$$10^{13}~\mathrm{cm}$, has been proposed as a possible binary companion to the progenitor \citep{van,maud}. Ref.
\cite{melina} also used hydrodynamical modeling to argue against a single-star progenitor scenario.

To better constrain the SN shock interaction with the CSM and the early evolutionary behavior, as well as the properties of the progenitor star, many authors analyzed the early X-ray emission from SN 2011dh. \citet{soderberg} analyzed \textit{Swift} observations from $\sim$3.5 to 30~days after the explosion, along with two \textit{Chandra} observations at approximately 11 and 32~days. They used a power-law model to fit the data. They reported an X-ray luminosity decreasing as t$^{-0.8 \pm 0.2}$, and derived a mass-loss rate of $\dot{M} \sim 6 \times 10^{-5}\,M_\odot\,\mathrm{yr}^{-1}$, with a wind velocity $v_w = 1000~\mathrm{km\,s^{-1}}$ assuming a Wolf--Rayet star progenitor.

\citet{2mxmm} analyzed two early \textit{XMM-Newton} observations using thermal, two-component models to fit the data. They interpreted the emission as being dominated by the reverse shock 10~days after the explosion. Ref. \cite{2012xmm}, on the other hand, fitted the spectra with a power-law component and found a good fit. They needed a harder spectral component during the first 10 days and a softer component at later times. They interpreted the late-time power-law component as either inverse Compton emission from radio synchrotron emitting electrons or emission from the reverse shock. 

\citet{maeda} extended the analysis of X-ray observations to $\sim$500~days after the explosion, reporting a similar power-law index to \citet{soderberg} but with a mass-loss rate an order of magnitude higher than that obtained by \citet{soderberg}: $\dot{M} \sim 3 \times 10^{-6}\,M_\odot\,\mathrm{yr}^{-1}$ assuming $v_w = 20~\mathrm{km\,s^{-1}}$. {For a wind velocity of $20~\mathrm{km\,s^{-1}}$, the mass-loss rate of Soderbergh et al. would become $\dot{M} \sim 1.2 \times 10^{-6}\,M_\odot\,\mathrm{yr}^{-1}$. The analysis of \citet{maeda} }differed from the one by \citet{soderberg} in that they fitted the full spectrum with thermal rather than power-law models, arguing that non-thermal emission would not persist over $\sim$500~days. They concluded that the X-ray emission of SN~2011dh originated from an adiabatic reverse shock.

In this paper, we extend the X-ray light curve of SN 2011dh up to the present day, given the large amount of archival data present. In Section 2
 we describe the X-ray observations from various satellites and how they were analyzed. Section 3 
 shows the results for the X-ray lightcurve and derives the mass-loss rate from the lightcurve under various assumptions. Section 4  
 discusses our results in the context of SN 1993J. Our conclusions are reported in Section 5.

\section{Observations}
SN 2011dh has been observed by most major X-ray satellites, including \textit{XMM, Chandra} and \textit{Swift}. We have processed all available observations of SN 2011dh taken with the \textit{Chandra} satellite, as well as the two earliest observations with the \textit{XMM-Newton} satellite. In order to extend the lightcurve over as large a time baseline as possible, we also analyzed all the \textit{Swift} observations taken in 2025. In addition we have included the early \textit{Swift} observations up to 10 days that were analyzed by other authors. This has allowed us to trace the evolution of SN 2011dh over a fourteen-year period following its explosion on {UT 2011 May 31}. Basic information regarding the observations analyzed by us is summarized in {Table \ref{tab:obs}}.

\begin{table}[H]
\caption{List of all observations.\label{tab:obs}}
\begin{adjustwidth}{-\extralength}{0cm}
\begin{tabularx}{\fulllength}{lCCCC}
\toprule
\textbf{Observation ID} & \textbf{Days After Explosion {*}} & \textbf{Instrument} & \textbf{Count Rate (cts s$^{-1}$)} & \textbf{Exposure Time (ks)} \\
\midrule
0677980701 & 6.21 & XMM PN & $(6.19 \pm 0.51)\times 10^{-2}$ & 4.3 \\
0677980801 & 10.20 & XMM PN & $(4.48 \pm 0.68)\times 10^{-2}$ & 2.0 \\
12562
 & 11.29 & Chandra ACIS-S & $(1.26 \pm 0.13)\times 10^{-2}$ & 9.6 \\
12668 & 32.44 & Chandra ACIS-S & $(4.71 \pm 0.81)\times 10^{-3}$ & 9.9 \\
(13812, 13813) & 468.25 & Chandra ACIS-S & $(1.56 \pm 0.09)\times 10^{-3}$ & 358.4 \\
(15496, 13814, 13815, & 486.71 & Chandra ACIS-S & $(1.45 \pm 0.08)\times 10^{-3}$ & 408.7 \\
13816, 15553) &&&&\\
19522 & 2116.03 & Chandra ACIS-I & $(1.32 \pm 1.21)\times 10^{-4}$ & 37.8 \\
20998 & 2648.67 & Chandra ACIS-S & $(8.58 \pm 3.23)\times 10^{-4}$ & 19.8 \\
(23472--23476) & 3491.98 & Chandra ACIS-S & $(2.11 \pm 1.22)\times 10^{-4}$ & 173.2 \\
(23477--23481, 25689) & 3662.52 & Chandra ACIS-S & $(1.91 \pm 1.12)\times 10^{-4}$ & 167.3 \\
(00011417207--8, \textsuperscript{a} & 5084.28 & Swift XRT & $(1.10 \pm 0.18)\times 10^{-3}$ & 33.7 \\
00011417210--18 &  &  &  &  \\
00011417220--29, &  &  &  &  \\
00011417231) &  &  &  &  \\
\bottomrule
\end{tabularx}
\end{adjustwidth}
\noindent{\footnotesize {Observations within parentheses are grouped together}. Count rates are calculated using \textit{Xspec 12.14.1}.
{*} {The time of explosion in JD is 2455713.393.}
\textsuperscript{a} Source count rate. Source and background were fitted simultaneously.}
\end{table}

\subsection{Chandra}

All Chandra observations have been processed using CIAO 4.16 and CALDB 4.11.3. The SN is assumed to be centered on the following coordinates: RA 13:30:05.1055 and DEC +47:10:10.9227. The spectrum for each observation is generated using the \textit{specextract} function. A circular region with a radius of 4$^{\prime \prime}$ is used to extract the source spectrum, while a similarly sized region some distance away is used for the background spectrum. Spectral analysis is carried out with the Sherpa version included with CIAO 4.16.

Seven \textit{Chandra} observations were taken in 2012, all within a short time interval. The first two, taken 3 days apart, had a total exposure of about 337 ks and were combined together. The remaining 5, spanning 3 weeks, were also grouped together, resulting in a total exposure time $>$ 400 ks. Eleven observations were taken in 2020 and 2021, spaced roughly one month apart. These were combined together in 2 groups, leading to an increase in the total number of counts, and thus enabling more reliable fitting. The groupings were justified, given that the light travel time across the SN at that epoch would be smaller than the time period over which the observations were taken.

\subsubsection*{Fitting}
We used \textit{Sherpa} to fit the spectrum, adopting the \textit{vapec} thermal model. A redshift of 0.0015 was adopted for the SN. {The \textit{chisquared-gehrels} statistic was used as the fit statistic in most cases. In those cases where the source and background data were fitted separately, we used the \textit{cstat} statistic to find the best fit. Moreover, 1-$\sigma$ confidence errors were found for each quantity using the Sherpa \textit{covar} routine. This routine computes confidence interval bounds for the specified model parameters in the dataset, using the covariance matrix of the statistic. Errors on the absorbed and unabsorbed flux were computed using the \textit{sample\_flux} routine in Sherpa. For each iteration, this routine draws the parameter values of the model from a normal distribution, filters out samples that lie outside the soft limits of the parameters, evaluates the model, and sums the model over the given range to return the flux.}

Initially, the column density $N_H$ and electron temperature $kT$ were allowed to remain as free parameters.  The best-fit values from the spectral fitting are given in {Table \ref{tab:flux}}. Up to approximately 500 days after explosion, the fitted $N_H$ value is higher than the Galactic column density, \(0.0154 \times 10^{22}~\text{cm}^{-2}\) \citep{dl90}. Values of \textbf{$N_H$} exceeding the Galactic value were also found by \citet{2mxmm}. At epochs exceeding 500 days, the $N_H$ value was found to be equal to or less than the Galactic value, so we adopted the Galactic value at those epochs.

An $N_H$ value of $0.018 \times 10^{22}~\text{cm}^{-2}$ was adopted by \cite{soderberg,maeda}. We therefore also tried this value of the column density $N_H$ for observations up to 500 days after the explosion. For observations at 6, 10, and 11 days, this had a minor impact on the fitted $kT$ values and the total unabsorbed flux. However, at 32 days, it resulted in a larger modification of the flux density at 1~keV ( discussed in more detail in Section 4.)

\citet{soderberg} adopted a \textit{power-law} model for fitting the observations, while ref.
~\cite{2mxmm} adopted a two-component thermal model (\textit{mekal} + \textit{mekal}) for observations at a similar epoch. Ref. \cite{maeda} used a thermal model to fit the data at $\sim$500 days. Given these contrasting results, we tested both the \textit{power-law} and \textit{vapec} models and obtained consistent results for observations up to approximately 10 days after the explosion (including the XMM data described in Section~\ref{sec3.2}). For the Chandra observation at 11 days, {using the \textit{vapec} model} we obtained an unabsorbed flux of \(11.76^{+3.16}_{-3.46} \times 10^{-14}~\text{erg}~\text{cm}^{-2}~\text{s}^{-1}\), which is consistent with the best-fit value of \(10.0 \times 10^{-14}~\text{erg}~\text{cm}^{-2}~\text{s}^{-1}\) \mbox{reported by \citet{soderberg}.}

\begin{figure} [H]
    \includegraphics[width=0.75\linewidth]{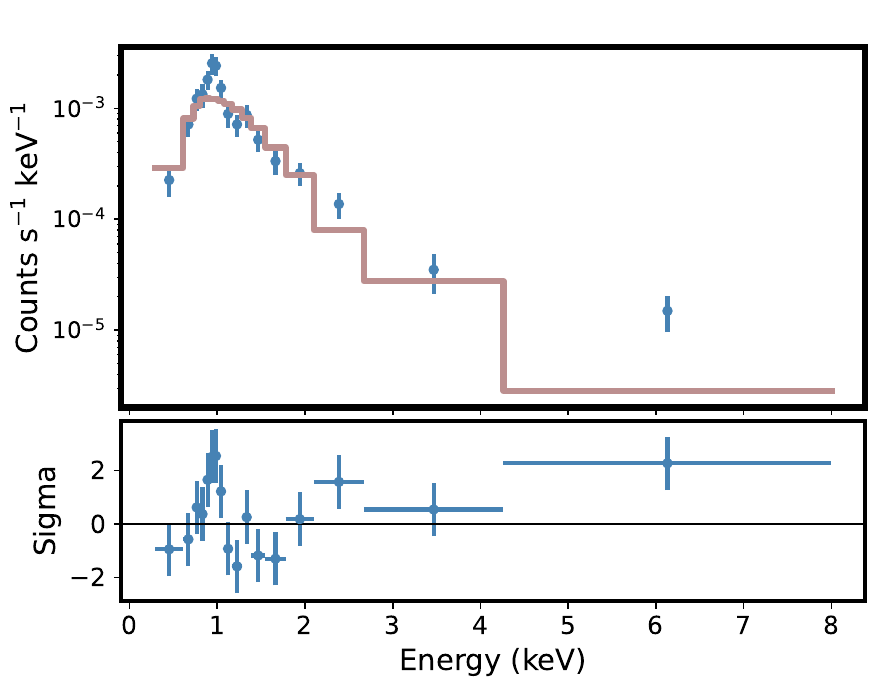}
\caption{The
 \textit{power-law} fit of the combined spectra of observations 13812 and 13813. Data points in dark blue, fit in brown.The fit completely misses the data point at 6.1 keV, and is inconsistent with the data at 1 and 2.6 keV.
\label{fig:powerlaw}}
\end{figure}

However, for later observations, we found that the \textit{power-law} model does not fit the spectrum at energies higher than 1 keV (see {Figure~\ref{fig:powerlaw}}). Additionally, for the observation at 32 days, we obtained an unabsorbed flux 2.4 times higher than that reported by \mbox{\citet{soderberg}.}

\renewcommand{\arraystretch}{1.2}
\begin{table}[H]
\caption{Flux from observations.\label{tab:flux}}
\begin{adjustwidth}{-\extralength}{0cm}
\begin{tabularx}{\fulllength}{lCCCC}
\toprule
\textbf{Observation ID} & \textbf{Days After Explosion {*}} & \boldmath{$N_H$} \textbf{(}\boldmath{$10^{22}$} \textbf{cm}\boldmath{$^{-2}$}\textbf{)} & \boldmath{$kT$}\textbf{ (keV)} & \textbf{Unabsorbed Flux (\boldmath{$10^{-14}$} erg cm\boldmath{$^{-2}$} s\boldmath{$^{-1}$})} \\
\midrule
0677980701 & 6.21 & 0.05 $^{+0.08}_{-0.04}$ & 56.24 $^{+12.21}_{-47.19}$ & 26.96 $^{+4.03}_{-4.74}$ \\
0677980801 & 10.20 & 0.016 $^{+0.08}_{-0.001}$ & 8.23 $^{+45.01}_{-4.74}$ & 13.68 $^{\pm 2.54}$ \\
12562 & 11.29 & 0.19 $^{\pm 0.17}$ & 3.44 $^{\pm 1.13}$ & 11.76 $^{+3.16}_{-3.46}$ \\
12668 & 32.44 & 0.67 $^{\pm 0.61}$ & 0.61 $^{+0.31}_{-0.46}$ & 6.56 $^{+4.84}_{-3.85}$ \\
13812, 13813 \textsuperscript{a} & 468.25 & 0.06 $^{\pm 0.04}$ & 0.98 $^{+0.06}_{-0.07}$ & 1.15 $^{+0.34}_{-0.27}$ \\
 &  &  & 11.7 $^{\pm 1.4}$ &  \\
15496, 13814, 13815 \textsuperscript{b} & 486.71 & 0.09 $^{+0.08}_{-0.06}$ & 0.90 $^{\pm 0.10}$ & 1.08 $^{+0.58}_{-0.41}$ \\
13816, 15553&  &  & 3.22 $^{+2.1}_{-1.7}$ &  \\
19522 & 2116.03 & 0.016 $^{+0.33}_{-0.001}$ & 0.77 $^{\pm 0.23}$ & 0.23 $^{\pm 0.11}$ \\
20998 & 2648.67 & 0.0154 \textsuperscript{c} & 0.70 $^{\pm 0.20}$ \textsuperscript{c} & 0.42 $^{+0.33}_{-0.22}$ \\
23472--23476 & 3491.98 & 0.0154 \textsuperscript{c} & 0.96 $^{\pm 0.66}$ & 0.32 $^{+0.14}_{-0.13}$ \\
23477--23481, 25689 & 3662.52 & 0.0154 \textsuperscript{c} & 1.25 $^{\pm 0.47}$ & 0.28 $^{+0.16}_{-0.12}$ \\
00011417207--8, 00011417210--18 & 5084.28 & 0.0154 \textsuperscript{c} & 0.74 $^{\pm 0.20}$ \textsuperscript{c} & 0.57 $^{+0.37}_{-0.33}$ \\
00011417220--29, 00011417231 &  &  &  &  \\
\bottomrule
\end{tabularx}
\end{adjustwidth}
\noindent{\footnotesize
{*} {The time of explosion in JD is 2455713.393.}
\textsuperscript{a} Thawed Fe (both components), Mg (higher-$kT$ component), and S (lower-$kT$ component).
\textsuperscript{b} Thawed Fe (both components) and S (lower-$kT$ component).
\textsuperscript{c} Fixed value.
}
\end{table}

\vspace{-0.8cm}

\begin{figure}[H]
{(a)}\includegraphics[width=0.47\linewidth]{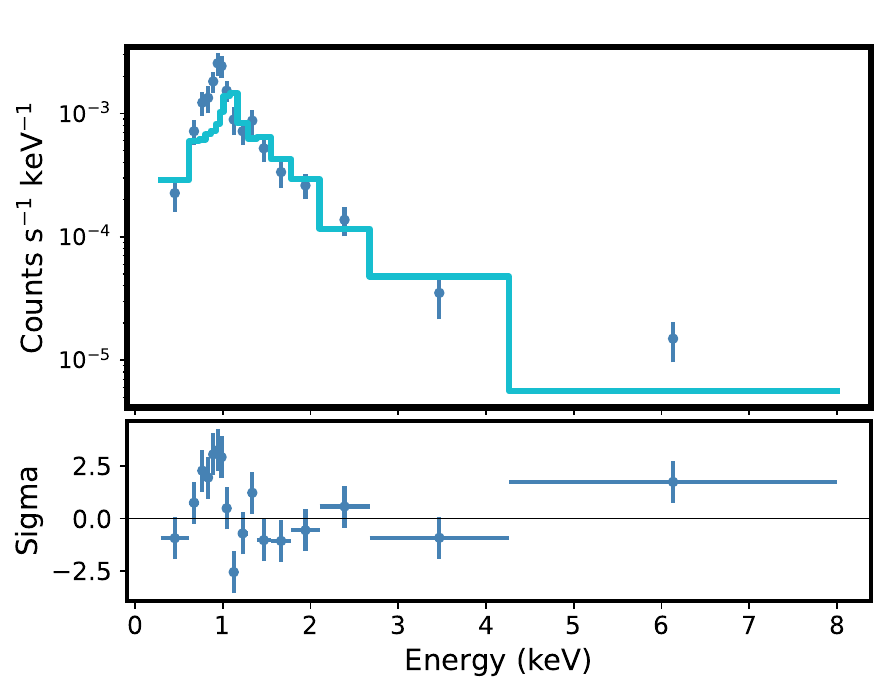}{(b)}\includegraphics[width=0.47\linewidth]  
{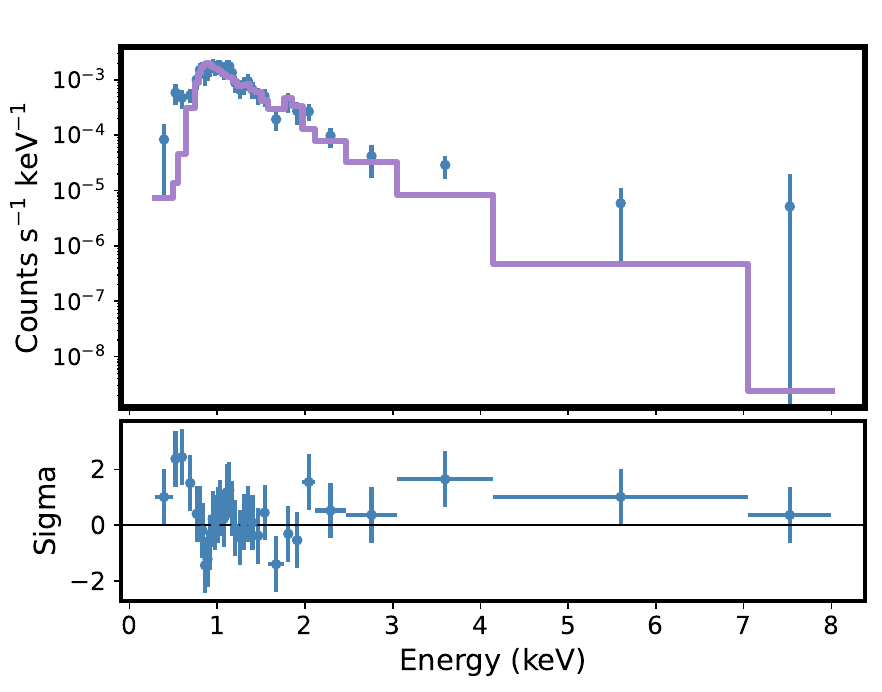}
{(c)}\includegraphics[width=0.47\linewidth]{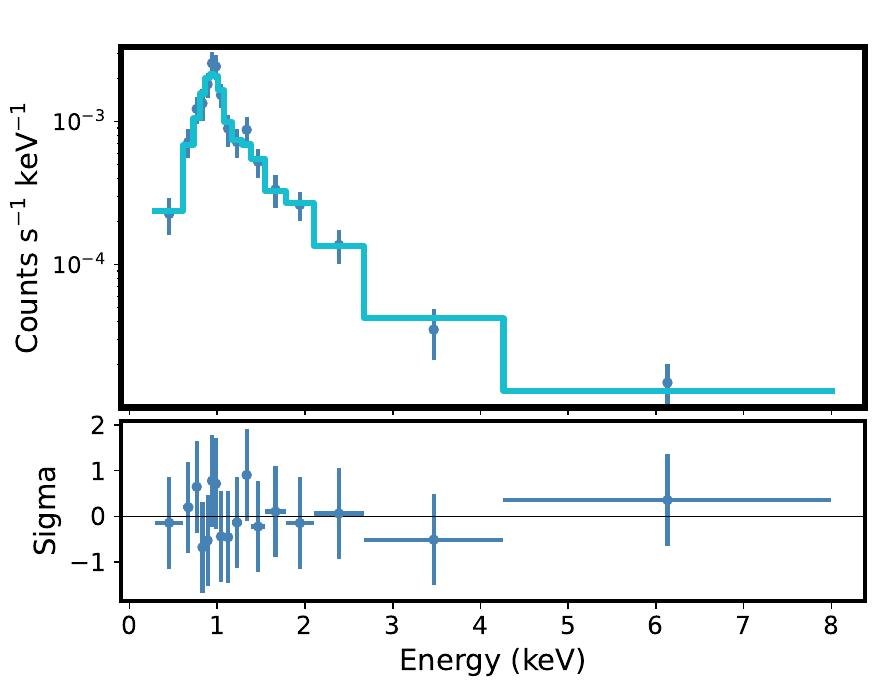}{(d)}\includegraphics[width=0.47\linewidth]{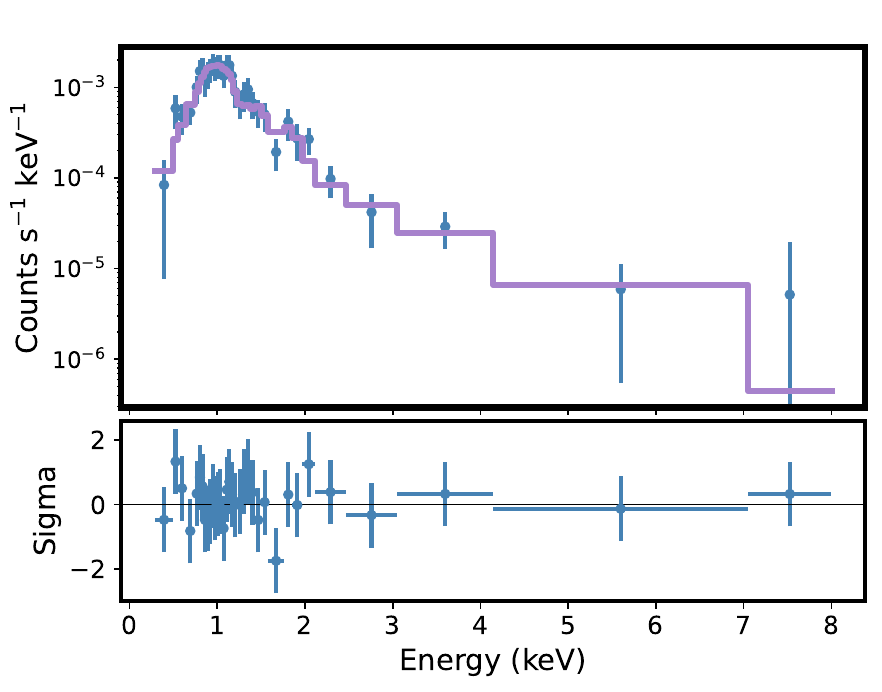}
\caption{{The combined spectra and initial fit of observations 13812 and 13813 is reflected by (a), with the improved fit reflected by (c). Data points in dark blue, fits in light blue. Similarly, the combined spectra and initial fit of observations 15496, 13814, 13815, 13816, and 15553 are reflected by (b), with improved fit reflected by (d). Data points in dark blue, fits in purple}
\label{fig:fits}}
\end{figure}

Using the \textit{vapec} model with all element abundances frozen at the solar value (see {Figure~\ref{fig:fits}}a,b) does not result in a good fit at 468 days. Thawing the element Fe improved the fit around 1 keV. We tested thawing other elements: Ne, S, Ar, C, Mg, and combinations thereof. We also introduced a two-component plasma thermal model (\textit{vapec} + \textit{vapec}) with different combinations of elements thawed. The two-component model provides the best fit. The best-fit model parameters are presented in {Table \ref{tab:flux}}, and the improved fits are shown in  {Figure~\ref{fig:fits}}c,d. The maximum likelihood ratio test shows that the two-component model is favored over any single-component model with about 99\% confidence.

A two-component model can be generally interpreted as X-ray emission arising from both the SN forward and reverse shocks. It is also possible that both components arise from an asymmetric reverse shock that encounters different levels of ejecta material. The higher-temperature component ($kT$ = 11.7 $\pm$ 1.4 keV) is enriched in Fe (8.37 $\pm$ 7.07~$Z_\odot$) but contributes only about 20\% of the total X-ray flux. The lower-temperature component ($kT$ = $0.98^{+0.06}_{-0.07}$ keV) has a super-solar S abundance (3.46 $\pm$ 1.8 $Z_\odot$), but a sub-solar Fe abundance (0.39 $\pm$ 0.17~$Z_\odot$). Given that both components are metal-rich, one in S and one in Fe, it is possible that both components could be arising from an aspherical reverse shock interacting with different layers of the metal-rich ejecta. 

For all observations {other than day 468}, the lack of sufficient counts meant that we had to rely on the use of a single-component \textit{vapec} model. It is possible that a second component may also exist at other epochs. Unfortunately, the low counts do not allow us to fit this component, and thus it remains unresolved. Given the fact that the lower temperature component contributed 80\% of the flux at day 468 and that the temperature is expected to decrease with time, we do not expect there to be a significant difference in flux  due to lack of a higher temperature component.

At 32 days, ref. 
 \cite{maeda} derived a flux $\approx$ 4 $\times 10^{-14} ~\text{erg}~\text{cm}^{-2}$ $\text{s}^{-1}$. Combining all observations between 467 and 498 days, they computed an X-ray flux of $0.7 \times 10^{-14}~\text{erg}~\text{cm}^{-2}$ $\text{s}^{-1}$. Our unabsorbed fluxes at 32, 468, and 486 days are somewhat higher than the best-fit values reported by them, although they are consistent within the error bars. We note that we allowed $N_H$ to vary in our fits, whereas \citet{maeda} fixed $N_H$ to their Galactic value, which is lower than the best-fit $N_H$ values we obtain at similar epochs (see Table~\ref{tab:flux}).

The Chandra observations with ObsIDs 19522 and 20998 have low counts. In this case, instead of subtracting the background counts from the source, we fit the background simultaneously with the source using a \textit{black-body} model. We find that the best-fit $N_H$ value is at or lower than the Galactic level of $0.0154 \times 10^{-22}$ cm$^{-2}$. We therefore fix the column density from this epoch onwards to the Galactic value.  Due to the low counts, we also fixed the value of $kT$ used to analyse observation 20998. As a further check, we used PIMMS v4.14, with the count rate and adopted $kT$ value to compute the flux, deriving unabsorbed fluxes consistent with those from the fit.

\subsection{XMM-Newton}

For the two \textit{XMM-Newton} observations at about 7 and 11 days, we reduced data from the MOS1, MOS2, and PN cameras using XMM-SAS 21.0.0. We selected circular regions with a 30$^{\prime\prime}$  radius for the source as well as the background, and limited the energy range to 0.5–10 keV. The counts from EPIC-MOS1 and EPIC-MOS2 were lower than those from the EPIC-PN camera. For the observation with ObsID 0677980701, the counts from PN, MOS1, and MOS2 were 378, 89, and 167, respectively; and for observation 0677980801, they were 144, 25, and 22.

Using \textit{Sherpa} and the \textit{vapec} model, and following the same fitting procedure as for the \textit{Chandra} spectra, we found that fitting all three spectra simultaneously yielded a flux nearly identical to that obtained from the PN spectrum alone. Using the latter alone returned smaller error bars. Therefore, we report the result from the PN spectra only (first two lines of Table \ref{tab:flux}). These are in agreement with the fluxes reported by \cite{2012xmm}.

\citet{2mxmm} analyzed the two XMM observations using a two-component \textit{MEKAL} + \textit{MEKAL} model. They obtained an ($N_H$) value of \(4.7^{+5.5}_{-3.3} \times 10^{20}~\text{cm}^{-2}\), along with unabsorbed fluxes of \(2.4\) and \(1.1 \times 10^{-13}~\text{erg}~\text{cm}^{-2}~\text{s}^{-1}\), respectively. The flux values we obtain are consistent with theirs. They are also comparable to those reported by
~\cite{2012xmm,maeda}.

\subsection{Swift}
In order to extend the X-ray light curve to the present day, we analyzed all \textit{Swift} observations taken in 2025. Ref.
 {\cite{soderberg} had analyzed all Swift observations up to day 30. We did not analyze \textit{Swift} observations after day 30 and before the year 2025, since the counts are low, and we already had extensive coverage from higher resolution observations for about 4000 days.} The {\textit{Swift} spectra} were extracted using the online Swift Data Analysis Tool at the University of Leicester ({Available
 at \url{https://www.swift.ac.uk/user\_objects/}, accessed on 10 September 2025}), which runs on HEASOFT v6.35.2. The spectra were also combined into one single spectrum. Due to the low count rate, we fixed the $N_H$ value to the Galactic value, and set the best-fit $kT$ to 0.74 keV, consistent with the $kT$ values from previous \textit{Chandra} and \textit{XMM} observations (see Figure~\ref{fig:evol}).  We first fit the background counts with a \textit{blackbody} model and subsequently fit the source and background together in a similar manner to all the other observations described previously.

\section{Results}

\subsection{{Time Evolution of Unabsorbed Flux, Temperature and Column Density}}

\begin{figure}[H]
\centering
\includegraphics[width=1\linewidth]{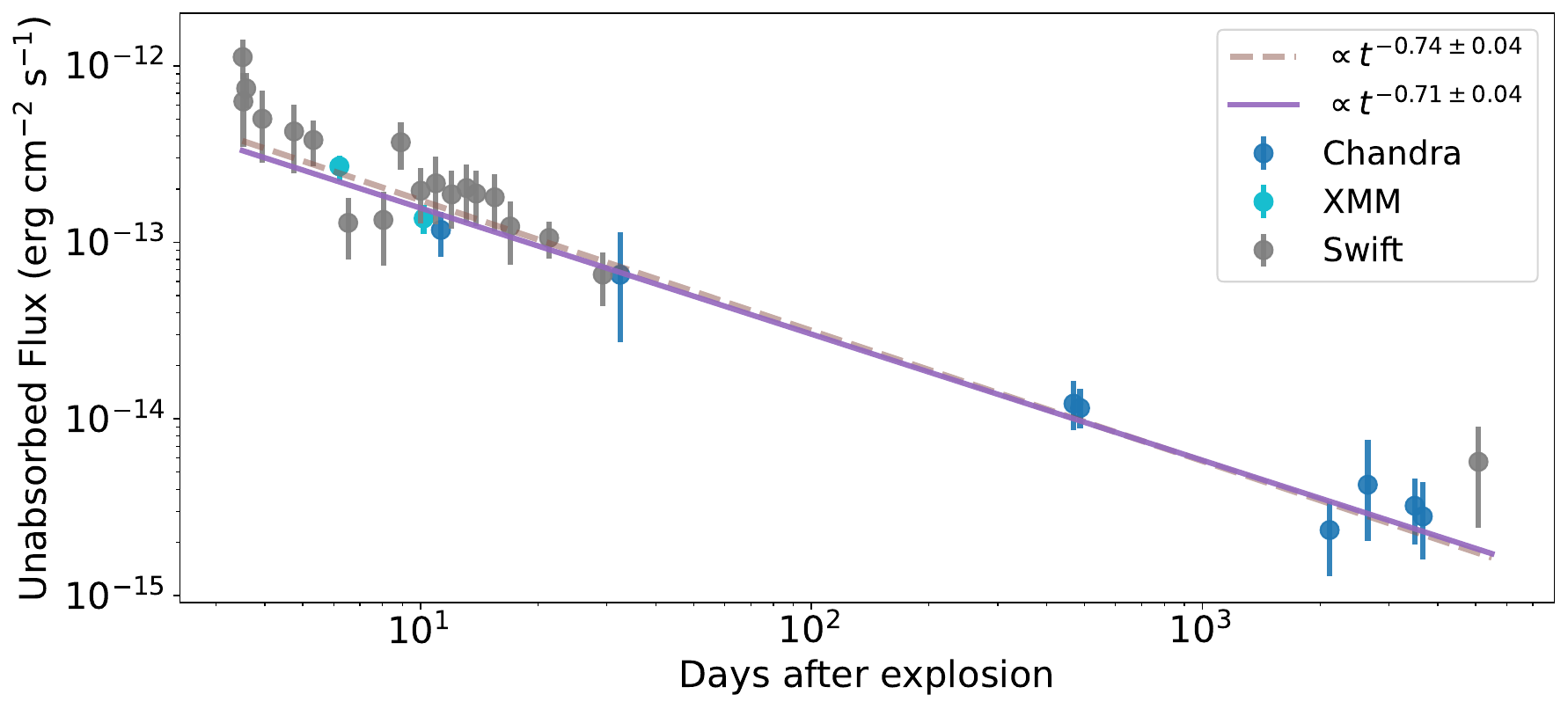}
\includegraphics[width=0.49\linewidth]
{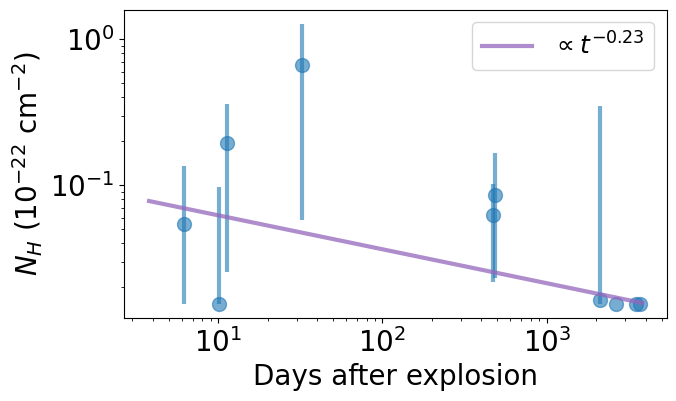}
\includegraphics[width=0.485\linewidth]
{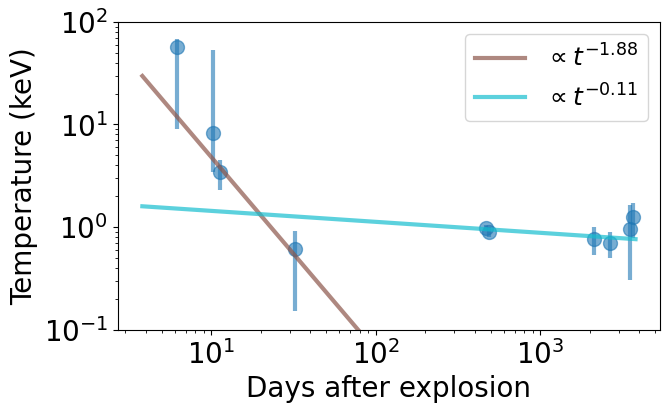}
\caption{Unabsorbed flux, Column Density $N_H$, and Temperature ($kT$) over time. {The $N_H$ and $kT$ values are extracted from the \textit{XMM} and \textit{Chandra} observations.}
\label{fig:evol}}
\end{figure}

In Figure~\ref{fig:evol}, we show plots of the unabsorbed flux, temperature $kT$, and column density $N_H$ as a function of elapsed time since the SN explosion. Assuming that the quantities decay as a power-law with time, we calculate the power-law index in each case and overlay the best-fit power-law model on each plot. The power-law decay index is provided in the plot legend for each figure.

\renewcommand{\arraystretch}{1.0}

{In order to find the line of best-fit to the flux values, column density, and temperature evolution with time in Figure~\ref{fig:evol}, we used the \textit{curve-fit} function from the SciPy open source library. This uses a non-linear least squares method to compute the best fit to the data.  }

{We have been very careful when comparing observations from different missions. We tried to capture as much of the flux from the SN as possible. For the Chandra observation, a 4$^{\prime\prime}$ region encircles almost 90\% of the emitted flux for a point source object located at or near the aim-point. In the case of XMM-Newton, the 30$^{\prime\prime}$ region encloses between 83 and 88\% of the flux, depending on the telescope used. However, the software compensates for missing flux. Since the Swift spectra had low counts, we combined them all into a single spectrum. We used the Sherpa software to fit all the data from the various satellites, to ensure that no errors crept in due to differences in software. We tried different techniques where possible, such as both fitting the data and using the PIMMS software to compute the flux in some cases. And we have tried to be consistent with the various quantities, such as fixing the column density to the Galactic value for all observations after about 2000 days. Through all this, we have attempted to minimize errors that may arise when comparing data from different telescopes and instruments.}

For the {evolution of the} unabsorbed flux, we carried out two power-law fits. The magenta line shows the fit for all observations from \textit{Chandra}, \textit{XMM-Newton}, and \textit{Swift} that we processed. In the second case, we also added the values of the flux from the \textit{Swift} observations analyzed by \cite{soderberg}. This fit is shown by a yellow line. The best-fit values of the power-law exponent in each case are consistent with each other within the error bars. They are also consistent with the result from \cite{maeda}, who found that the flux decays as t$^{-0.8 \pm 0.2}$. Thus, the luminosity decrease, which had been calculated up to 500 days, appears to continue with the same slope up to approximately 5100 days.

 The temperature $kT$ shows a clear decreasing trend over the first four epochs, or over approximately 32 days after the explosion. {It should be noted that the observation at 32 days had low counts, and the temperature is not well constrained}. In the subsequent observations, the temperature remains relatively stable, with a value fluctuating around 1 keV. Therefore, we fitted the data with two separate power-laws: one for the first four observations and another for the remaining data. The resulting power-law exponents are $-1.88$ for the early phase and $-0.11$ for the later phase. At late phases after 32 days, the temperature appears relatively stable.

The column density $N_H$ also shows a decreasing trend overall, with the exception of the data point at 500 days. At late epochs, $N_H$ approaches the Galactic column density of $0.0154 \times 10^{-22}~\mathrm{cm}^{-2}$, reaching this value around 2000 days after the explosion. After this epoch, $N_H$ is fixed at the Galactic value.

\subsection{Determination of the CSM density}\label{sec3.2}
In order to determine the evolution of the CSM density, we use the self-similar model proposed by \cite{1982}
for the evolution of the SN ejecta with time {(see also \cite{cf94})}. The SN density is assumed to evolve as ${\rho}_{SN} \propto v^{-n}$, whereas the CSM density goes as ${\rho}_{CSM} \propto r^{-s}$. A steady wind with constant parameters (mass-loss rate and wind velocity) would have $s=2$, whereas $s=0$ denotes a constant density medium. Other values of $s < 3$ would imply a wind where the parameters are changing with time. {The requirement that the energy be finite constrains the parameter $n$ to be $>$5. Its value is generally taken to be somewhere in the range 9--12~\cite{mm99}, although a steeper value is not restricted, and has sometimes been assumed \cite{maeda}.}

To extract the value of $n$ and $s$ for SN 2011dh, we  use the prescription from \citep{franssonetal96, dg12, Ramakrishnan_2020} for the evolution of the X-ray luminosity with time in the case of a non-radiative shock. We make the assumption here that the shock is non-radiative, in anticipation of the results that will be derived later on. In \citep{Ramakrishnan_2020} the X-ray luminosity in this model can be shown to evolve~as

\begin{equation}
    L_x \propto t^{\frac{-(12-7s+2ns-3n)}{n-s}}\\
     \propto t^{\frac{n-3}{n-s}(4-2s)-1}
\end{equation}

Plugging in the exponent $-0.74 \pm 0.04 = \frac{n-3}{n-s}(4-2s)-1$ with $n$ ranging from 10--20, we got $s$ $\approx$ $1.85 \pm 0.02$. The mass-loss from the progenitor star is not a steady wind with constant mass-loss parameters, although it is close to a steady wind, which would have $s=2$.

The temperature at the forward shock can be calculated using the Rankine-Hugoniot shock-jump conditions:

\begin{equation}
\label{eqn:temp}
T_{\mathrm{FS}} = 2.27 \times 10^9\, \mu_s \left(\frac{V_{\mathrm{FS}}}{10{,}000\, \mathrm{km\,s}^{-1}}\right)^2\, \mathrm{K}
\end{equation}

The temperature depends on the shock velocity, which can be obtained from the shock radius.  \citet{Bietenholz_2012} reported VLBI observations of SN~2011dh. By combining their measurements with those derived from synchrotron self-absorption (SSA) analysis, they provided a prescription for the evolution of  the radius of SN~2011dh evolving with~time:

\begin{equation}
\label{eq:radius}
R = 5.85 \times 10^{15} \left( \frac{t}{30~\text{days}} \right)^{0.92 \pm 0.03}~\text{cm}.
\end{equation}

Taking the derivative of this equation with time provides the forward shock velocity. At 500 days, this gives $V_{\mathrm{FS}} = 16{,}600 \pm 500\, \mathrm{km\,s}^{-1}$ . Plugging this value for the velocity into Equation \eqref{eqn:temp}, and taking  $\mu_s = 0.61$ for solar metallicity we obtain:

\begin{equation}
T_{\mathrm{FS}} \approx 3.8^{+ 0.25}_{- 0.2} \times 10^9\, \mathrm{K}
\end{equation}

This is significantly higher than the observed temperature of $kT \approx 1$~keV for {observations at 468 and 487 days. The shock velocity goes as 
 $t^{-0.08}$ which means that the temperature drops as $t^{-0.16}$. This is a slow decrease, and the post-shock temperature suggested by Equation~\eqref{eqn:temp} is always considerably higher than the measured X-ray temperature. This suggests that the X-ray emission {after 32 days arises from a lower temperature region, presumably the reverse shock, rather than the forward shock}. This is consistent with the results of \cite{maeda}, who argued for a reverse-shock origin of the X-ray emission at 500 days. 

In the context of the  self-similar solution derived by \cite{1982}, we can relate the temperature of the reverse shock to that of the forward shock \cite{Chevalier2003}:

\begin{equation}
\label{eqn:shktemp}
T_{\mathrm{RS}} = \left(\frac{3 - s}{n - 3}\right)^2 T_{\mathrm{FS}}
\end{equation}

Using   $s=1.85$ we can infer that $n$ should lie between 23 and 25. This value is somewhat higher than that in \cite{maeda}, because we have adopted a value of $s = 1.85$ rather than the value of $s = 2$ adopted by \cite{maeda}. Since both the forward shock velocity $V_{\mathrm{FS}}$ and the reverse shock temperature $T_{\mathrm{RS}}$ show only a small variation in later observations, we adopt $n = 24$ for all subsequent analyses. Applying the procedure to earlier observations, we find $n \sim 13$ at day~6, $\sim$17 at day~10, and $\sim$ 24 from day~32 onward. 

The above equation assumes that the electron and ion temperatures are in equilibrium, which is generally not true for young SNe. Ref. \cite{maeda} showed that electron-ion equilibration was  true for SN 2011dh up to 500 days. Even if the electron and ion temperatures are not in equilibrium, they are expected to equilibrate over time due to Coulomb collisions and plasma processes. Thus, if they were in equilibrium before 500 days, we would expect that they would more than likely be in equilibrium at later times. Or the deviations will be small. Therefore, we adopt the value $n=24$ for all epochs after 32 days. This value of $n$
 is large, but not unusually so. Ref. \cite{nadezhin85} in their derivation of the self-similar solution discussed the properties of SN ejecta with $n=20$, and Ref. \cite{cf94} have discussed the properties of SNe with steep ejecta density gradients of $n=20$, $n=50$, and even $n=\infty$.

The above derivation of the value of $n$ is only possible in the context of the
 \cite{1982,nadezhin85} self-similar solution, which is used to relate the forward and reverse shock temperatures. This description is standard and is used by numerous authors in the literature to describe young SN evolution \cite{franssonetal96, ncf09, chandraetal12b, dwarkadas14, maeda, drrb16, cf17} since detailed knowledge of the SN and CSM properties of the point-source extragalactic object is generally unavailable.

The value of $n$ for the first three observations is lower than the value of $n=24$ for the remaining epochs. The cooling time and mass-loss rate at those values are uncertain. At the first two epochs, the temperature is unconstrained and may be significantly higher. This would imply that the X-ray emission could be arising from the forward shock or a combination of forward and reverse shocks. Ref.~\cite{soderberg} had fitted the emission in the first 30 days using a power-law model. While the power-law model is clearly not suitable after about 12 days, it is a possibility at the first epoch. The value of $n$ may also vary due to variations in the density behind the reverse shock. Thus, it is difficult to evaluate the value of $n$ in the first few days.

\textls[-15]{The temperature derived for the first few observations has large uncertainties—especially} the first, which approaches the upper limit of the model. We therefore also estimate $n$ from the time dependence of the radius, $R \propto t^{\frac{n-3}{n-s}}$. For this we use the value $\frac{n-3}{n-s} = 0.92$ reported by \cite{Bietenholz_2012}, who calculated it by fitting the radius up to $\sim 200$~days after the explosion. This yields $n \sim 16$ for $s = 1.85$ and $n \sim 14.5$ for $s = 2$, consistent with the estimate at day~10. If we use these $n$ values in Equation~\eqref{eqn:shktemp}, we find $T_{\mathrm{RS}} \approx$ 7~keV. This temperature lies at the lower bound of the error bar for the day~6 observation. Therefore, for the day~6 observation, the value of $n$ is somewhat uncertain. We test both $n$ values when calculating the mass-loss rate, as detailed in the next section. 

\subsection{Mass-Loss Rate}
Our derived value of $s=1.85$ indicates that the wind parameters {of the progenitor of SN 2011dh}, such as mass-loss rate and wind velocity, are changing with time or, interchangeably, radius. It is generally assumed (although without a firm basis) that it is the mass-loss rate {of the progenitor} that is changing with time \citep{franssonetal96,chandraetal12a, chandraetal12b,drrb16}. If so, the mass-loss rate {of the progenitor} must be calculated at a specific radius, and from that computed at any other radius assuming the value of $s$. In order to accomplish this, we use the formulation by \cite{franssonetal96}. 

If the shock is adiabatic, the luminosity of the reverse shock at 1 keV can be written as (Eq. 3.10 in \cite{franssonetal96})

\begin{equation}
\begin{aligned}
\label{eqn:adiab}
    L_{rev}(1 \space\text{keV}) 
    &\approx 1.74 \times 10^{37}\xi\frac{4(n-3)(n-4)^2}{(3-s)^2(4-s)^2}\\
    &\times T_{8}^{-0.24}e^{-0.0116/T_{8}}\space\left(\frac{\dot M_{-5}}{v_{w1}}\right)^2\space V_4^{3-2s}\\ 
    &\space \times \left(\frac{t}{11.57 \space \text{days}}\right)^{3-2s} \space \text{ergs} * \text{s}^{-1}\text{keV}^{-1}
\end{aligned}
\end{equation}

The total luminosity of a radiative reverse shock is given by Equation 3.17 in \citet{franssonetal96}

\begin{equation}
\begin{aligned}
\label{eqn:rad}
    L_{rev} = 1.57 \times 10^{41} \,\frac{2(3-s)^2(n-3)(n-4)}{(4-s)(n-s)^3} \, \left(\frac{\dot M_{-5}}{v_{w1}}\right) \times V_4^{5-s} \, \left(\frac{t}{11.57 \space \text{days}}\right)^{2-s}
\end{aligned}
\end{equation}

\noindent
where $\dot{M}_{-5}$ is the mass-loss rate in units of 10$^{-5}$ M$_{\odot}$ yr$^{-1}$, ${v_w}_1$ is the wind velocity in units of 10 km s$^{-1}$, $V_4$ is the maximum ejecta velocity in units of 10$^4$ km s$^{-1}$, and $t$ is the time in days. We take $\xi = 0.86$. The flux density at 1 keV is converted to luminosity density using a distance of 8.4 Mpc. We take the wind velocity to be $v_w$ = 10 km/s, and thus $v_{w1} = 1$.

Based on Equation~\eqref{eq:radius}, the supernova reached a radius of \(1.3 \times 10^{15}~\text{cm}\) with an average velocity of approximately \(2.36 \times 10^4~\text{km}~\text{s}^{-1}\) over the first six days. Therefore, \(V_{4,0}\) is taken to be 2.36 on day 6.

The shock velocity should be decreasing over time. We model the velocity  \(V_{4,t}\) at later times by taking the time derivative of $R$ from Equation~\eqref{eq:radius}, which gives the following:

\begin{equation}
    V_{4,t} = V_{4,0}*{(\frac{t}{t_0})}^{-0.08}
\end{equation}

As shown above, the emission must arise from the reverse shock. It remains to be decided whether the shock is adiabatic or radiative. In order to determine this, we calculate the mass-loss rate for both the adiabatic and radiative cases from Equations~\eqref{eqn:adiab} and \eqref{eqn:rad}. We then  use that mass-loss rate to compute the cooling time, using equations 3.14 and 3.7 from \citet{franssonetal96}.

For $kT < 2 \times 10^7~\mathrm{K}$ (applicable to observations taken 32 days and after),

\begin{equation}
\begin{aligned}
t_{\mathrm{cool}} = &\ 3.5 \times 10^9 \,
\frac{(4-s)(3-s)^{4.34}}{(n-3)(n-4)(n-s)^{3.34}} \\
& \times V_4^{\,3.34+s} 
\left( \frac{\dot{M}_{-5}}{v_{w1}} \right)^{-1}
\left( \frac{t}{11.57~\mathrm{days}} \right)^s \ \mathrm{s}
\end{aligned}
\end{equation}

For $kT > 2 \times 10^7~\mathrm{K}$ (applicable to observations between 6 and 11 days),

\begin{equation}
\begin{aligned}
t_{\mathrm{cool}} = &\ 1.6 \times 10^3 \,
\frac{(4-s)(3-s)^{2}}{2(n-s)(n-3)(n-4)} \\
& \times V_4^{\,5 - s} 
\left( \frac{\dot{M}_{-5}}{v_{w1}} \right)^{-1}
\left( \frac{t}{11.57~\mathrm{days}} \right)^{4-s} \ \mathrm{days}
\end{aligned}
\end{equation}

We inspect the cooling time to see if it is shorter or longer than the flow time (age of the supernova). A cooling time shorter than the flow time indicates a radiative shock. 

For all observations later than 32 days ($kT \sim 1 \space \text{keV}$ ), this exercise reveals that the cooling time is much larger than the age of the SN, and therefore that emission originates from an adiabatic reverse shock. 

In order to confirm {that the emission arises from the reverse shock}, we use the derived mass-loss rate in the luminosity equation given by \citet{Chevalier2003}:

\begin{equation}
L_i \approx 3.0 \times 10^{39} \, g_{\mathrm{ff}} \, C_n \left( \frac{\dot{M}_{-5}}{v_{w1}} \right)^2 \left( \frac{t}{10~\mathrm{days}} \right)^{-1} \ \mathrm{erg\ s}^{-1}
\end{equation}

\noindent
where we take $g_{\mathrm{ff}}$, the Gaunt factor, to be 1.  
Using $C_n = 1$ for the forward shock, we obtain luminosities of order {$\sim$$10^{35}-10^{37}~\mathrm{erg\ s}^{-1}$ over the entire time period}, which is two orders of magnitude lower than our measured luminosities.  

However, if we use the following in the above equation
\begin{equation}
\label{eq:rs}
C_n = \frac{(n-3)(n-4)^2}{4(n-2)}
\end{equation}

\noindent
as appropriate for the reverse shock, we find that $C_n \approx 95.5$ for the reverse shock with $n = 24$ and $C_n \approx 22.5$ for $n = 17$. This increases the shock luminosity to be in the range $\sim$$10^{37}$–$10^{39}~\mathrm{erg\ s}^{-1}$, consistent with the order of magnitude of our calculated values. This further confirms that the emission predominantly arises from the reverse shock for observations taken later than 10 days. 

It is possible that in the two-temperature fit, we have emission from different parts of the reverse shock, or from a forward and reverse shock. In the two-temperature fit for the $\sim$500-day observation, the lower-temperature component, around 1~keV, which likely originates from the reverse shock, has a normalization approximately four times larger than the high-temperature component and therefore dominates the emission.

For observations between 6 and 11 days, the $kT$ values obtained are significantly higher than those at later times. However, the magnitude of the emission fits the reverse shock better, and it would be difficult to obtain the same luminosity from the forward shock. Assuming an adiabatic reverse shock, we obtain cooling times larger than the epoch at which the observation was taken. Thus, the data are consistent with the assumption of emission arising from an adiabatic reverse shock for all observations.

For the observation at day\,6, we obtain a temperature of $kT \approx 56\ \mathrm{keV}$, with an upper limit of $\approx 68\ \mathrm{keV}$ and a lower bound near $9\ \mathrm{keV}$. As a result, the temperature is poorly constrained. We tested values of $n = 13$ and $n = 17$, corresponding to estimates based on the best-fit $kT$ and its lower limit. The $n = 13$ case yields cooling times and mass-loss rates consistent with emission originating from an adiabatic reverse shock, and the value is more consistent with the `{n'} 
 values from the other observations.

The mass-loss rate of {the progenitor of} SN~2011dh, calculated using all the observations, along with the corresponding cooling time, is presented in Table~\ref{tab:cooling}.

\begin{table}[H]
\renewcommand{\arraystretch}{1.2} 
\caption{Cooling times and mass-loss rates. \label{tab:cooling}}
\setlength{\tabcolsep}{26pt}
\begin{tabular}{ccc}
\hline
\textbf{Observation Time (Days)} & \boldmath{$t_{\mathrm{cool}}$} \textbf{(Days)} & \boldmath{$\dot{M}_{-5}$} \textbf{(}\boldmath{$M_\odot~\mathrm{yr}^{-1}$}\textbf{)} \\
\hline
6.21   & 39     & 0.22 \\
10.20  & 62      & 0.13 \\
11.29  & 77     & 0.13 \\
32.44  & 37      & 0.10 \\
468.25 & 1061   & 0.15 \\
486.71 & 1016  & 0.17 \\
2116.03 & 7104 & 0.20 \\
2648.67 & 12,017 & 0.16 \\
3491.98 & 17,271 & 0.17 \\
3662.52 & 27,828 & 0.12 \\
5084.28 & 44,641 & 0.11 \\
\hline
\end{tabular}
\renewcommand{\arraystretch}{1.0} 
\end{table}

\section{Comparison with Type IIb SN 1993J}

SN~1993J is perhaps the best-studied Type~IIb supernova so far, due to its proximity. In Figure~\ref{fig:flux} we plot the unabsorbed flux of SN~1993J from \cite{chandra2009}, alongside that of SN~2011dh (Figure \ref{fig:flux}). 

\begin{figure}[H]
\includegraphics[width=0.8\linewidth]{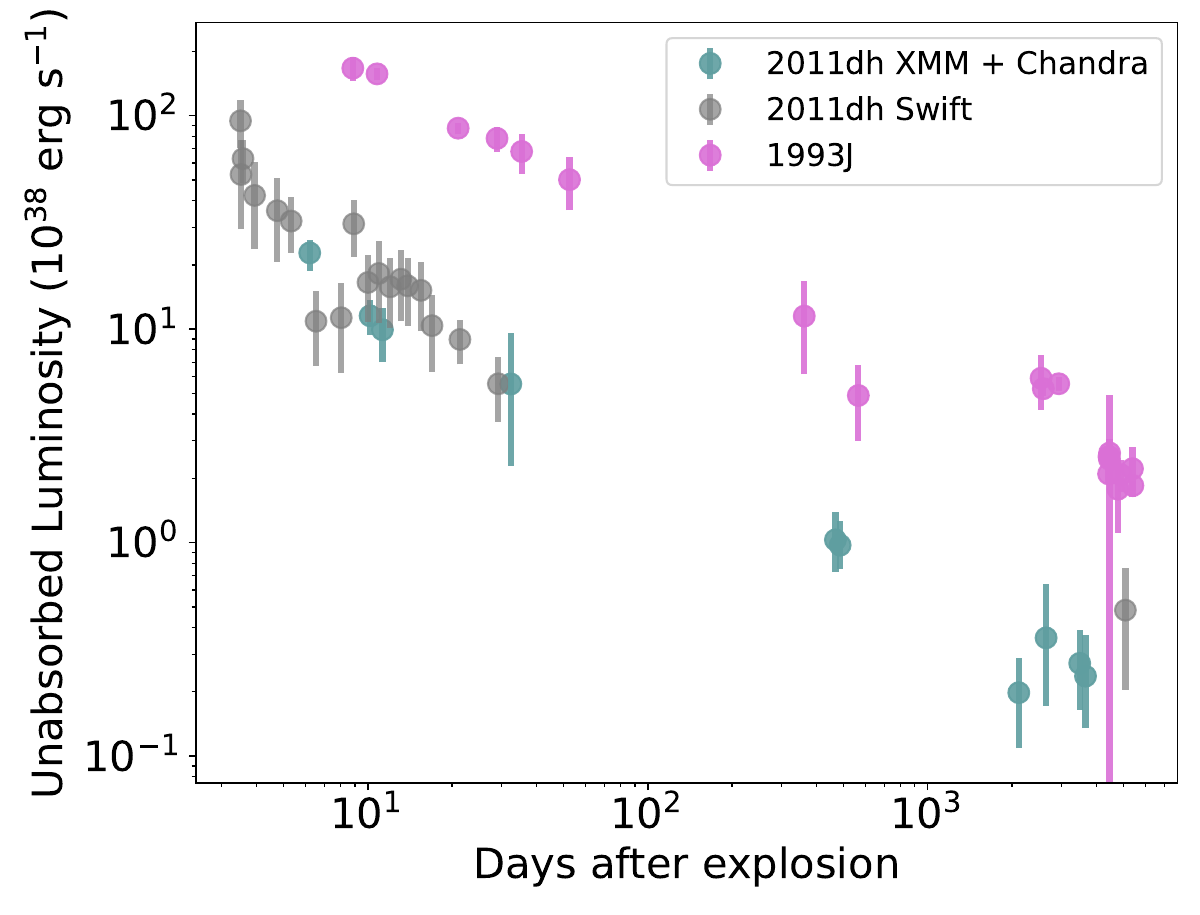}
\caption{Comparing the flux evolutions of SN 1993J and {SN 2011dh}.}
\label{fig:flux}
\end{figure}

Based on this comparison, the luminosity declines at a similar rate for both supernovae up to about 3500 days, with SN~1993J being consistently around an order of magnitude more luminous than SN~2011dh. Since the luminosity at a given radius is proportional to the square of the density, or alternatively the mass-loss rate (divided by velocity), we expect the mass-loss rate of {the progenitor of} SN 2011dh to be lower by factors of a  few.  Caveats to this are that the evolution of the CSM density may not be the same, the SN ejecta profile may not be the same, and/or the temperatures may not be the same.  The mass-loss rate of {the progenitor of} SN~1993J was estimated by \citep{franssonetal96} to be approximately \(4 \times 10^{-5}~M_\odot~\mathrm{yr}^{-1}\), based on observations during the first two weeks after the explosion, and \(3.8 \times 10^{-5}~M_\odot~\mathrm{yr}^{-1}\) reported by \cite{ta}, analyzing $\sim$8.5 years of data. This is about an order of magnitude larger than that of {the progenitor of} SN~2011dh. For a mass-loss rate an order of magnitude lower, we would expect the luminosity of SN 2011dh to be about two orders of magnitude lower if all other conditions are the same. The mass-loss rate of {the progenitor of} SN 2011dh is thus lower than would be expected from its luminosity, or alternatively, the luminosity is higher than expected for the corresponding mass-loss rate.  One of the major differences between SN1993J and SN 2011dh is that the value of $n$ is much higher in SN 2011dh than in SN 1993J. As deduced in this paper, as well as in \citep{maeda}, $n \approx 24$ for SN 2011dh. On the other hand, ref.  
 \citep{ta} assumed that $n\approx8$ for SN 1993J. As evident from Equation~\eqref{eq:rs}, a larger value of $n$ enhances the reverse shock emission for a given mass-loss rate. The reverse shock is expanding into a steeper decline and higher ejecta density. The X-ray emission from the reverse shock, which depends on the value of $n$, is thus much higher in SN 2011dh than in 1993J, leading to the higher luminosity for the given {progenitor} mass-loss rate. The luminosity comparison also appears to confirm the high value of $n$ computed for SN 2011dh.

\section{Discussion and Conclusions}

We have analyzed the X-ray emission from Type~IIb SN~2011dh over a period of 14 years since its explosion on 2011 May 31 UT \citep{ar}. Incorporating \textit{Swift} data from \citet{soderberg}, together with our analysis of \textit{XMM-Newton}, \textit{Chandra}, and \textit{Swift} observations, we have constructed the X-ray light curve of SN~2011dh over a time baseline of $\approx 5100$ days. 

In 5100 days the {the forward shock of the} supernova has reached a radius of $6.58 \times 10^{17}$~cm based on Equation~\eqref{eq:radius}. For a wind speed of $10~\mathrm{km~s^{-1}}$, the circumstellar medium ejected by the progenitor star has been traveling for roughly {21,000} years, meaning that we are tracing the star’s mass-loss history out to $\sim$21,000 years before explosion (or $\sim$10,500 years if a wind speed of $20~\mathrm{km~s^{-1}}$ is assumed).

Our light curve shows that the luminosity continues to decay with almost the same time dependence as that modeled by \citet{maeda} using X-ray data from the first 500 days. This decay time has persisted through $\sim$14~years of supernova evolution.

{Our derived progenitor mass-loss rate is $(1.0$--$2.2)\times10^{-6}~M_\odot~\mathrm{yr}^{-1}$, assuming $v_w = 10$ km s$^{-1}$, or $(2.0$--$4.4)$ $\times10^{-6}$ $M_\odot~\mathrm{yr}^{-1}$ for $v_w = 20~\mathrm{km~s^{-1}}$. These results are consistent with the estimate of $3\times10^{-6}~M_\odot~\mathrm{yr}^{-1}$ by \citet{maeda} for $v_w = 20~\mathrm{km~s^{-1}}$. 
The mass-loss rate at different epochs varies somewhat, but all values lie within a factor of 2.} We find that the X-ray emission from SN~2011dh is dominated by an adiabatic reverse shock. This interpretation is supported by calculating the cooling time at all epochs, all of which exceed the age of the corresponding observations. A similar conclusion was also reached by~\citet{2012xmm}, based on the two early \textit{XMM-Newton} observations, and \mbox{by~\citet{maeda},} who analyzed the supernova at 500~days, although the techniques used were different. Together, these results indicate that the luminosity of SN~2011dh has continued to decay in a remarkably persistent manner.

The mass-loss rate of {the progenitor of} SN 2011dh seems to be about an order of magnitude lower than that of {the progenitor of} SN 1993J. This would naively indicate a luminosity about two orders of magnitude lower, given that the luminosity is proportional to the square of the density or the {progenitor} mass-loss rate. However, the luminosity is only about an order of magnitude lower. We attribute this to the fact that the reverse shock is expanding in ejecta that are declining {in density} much more steeply {than SN 1993J}. The ejecta density in SN 2011dh decays with $n \approx 24$, whereas in SN 1993J $n\approx 8$. The higher value of $n$ leads to a much larger reverse shock luminosity in SN 2011dh.

\vspace*{+6pt}
\newpage
\authorcontributions{Methodology: E.J.G., V.V.D. Data curation: E.J.G.. Data visualization: E.J.G., V.V.D.  Writing original draft: E.J.G., V.V.D. All of the authors approved the final submitted draft. All of the authors have read and agreed to the published version of the manuscript.}

\funding{This research received no external funding. }

\institutionalreview{Not applicable.}

\dataavailability{All data is included in the tables and figures in the article.}

\acknowledgments{We thank the editor and referees for a very thorough reading of the paper, and for their most useful comments and suggestions, which have helped to substantially improve the paper.
Elisa Gao is grateful to have been awarded a Quad Fellowship from the University of Chicago during the academic year 2024--2025, as well as in the summer of 2025. She also acknowledges support from the NASA Space Grant in the summer of 2024 and the fall of 2025.
This research has made use of data obtained from the Chandra Data Archive provided by the Chandra X-ray Center (CXC). This research is also based on observations obtained with XMM-Newton, an ESA science mission with instruments and contributions directly funded by ESA Member States and NASA.}

\conflictsofinterest{The authors declare no conflicts of interest.} 
\begin{adjustwidth}{-\extralength}{0cm}
\reftitle{References}

\PublishersNote{}
\end{adjustwidth}

\end{document}